\documentstyle[epsfig,psfig]{elsart}
\begin{document}
\title{The primary cosmic ray composition between $10^{15}$ and
$10^{16}$ eV from Extensive Air Showers electromagnetic and TeV muon data}
%\author{}
\vskip 0.5cm
\bigskip
\begin{center}
\nobreak
\pretolerance=10000
{\bf The EAS-TOP Collaboration:}\\
\vskip 0.5cm M. Aglietta$^{19}$, B. Alessandro$^{16}$, P.
Antonioli$^{2}$, F. Arneodo$^{7}$, L. Bergamasco$^{16}$, M.
Bertaina$^{16}$, C. Castagnoli$^{19}$, A. Castellina$^{19}$, A.
Chiavassa$^{16}$, G. Cini$^{16}$
% ...
\\
\vspace{1.5 cm}
\footnotesize
1. Dipartimento di Fisica dell'Universit\`a  di Bari and INFN, 70126 Bari, 
 Italy \\
2. Dipartimento di Fisica dell'Universit\`a  di Bologna and INFN, 40126 
Bologna, Italy \\
3. Physics Department, Boston University, Boston, MA 02215, USA \\
4. California Institute of Technology, Pasadena, CA 91125, USA \\
5. Department of Physics, Drexel University, Philadelphia, PA 19104, USA \\
6. Laboratori Nazionali di Frascati dell'INFN, 00044 Frascati (Roma), Italy \\
7. Laboratori Nazionali del Gran Sasso dell'INFN, 67010 Assergi (L'Aquila), 
 Italy \\
8. Depts. of Physics and of Astronomy, Indiana University, Bloomington, IN 
47405, USA \\
9. Dipartimento di Fisica dell'Universit\`a  dell'Aquila and INFN, 67100 
L'Aquila, Italy\\
10. Dipartimento di Fisica dell'Universit\`a  di Lecce and INFN, 73100 Lecce, 
 Italy \\
11. Department of Physics, University of Michigan, Ann Arbor, MI 48109, USA \\
12. Dipartimento di Fisica dell'Universit\`a  di Napoli and INFN, 80125 
Napoli, Italy \\
13. Dipartimento di Fisica dell'Universit\`a  di Pisa and INFN, 56010 Pisa, 
Italy \\
14. Dipartimento di Fisica dell'Universit\`a  di Roma "La Sapienza" and 
INFN, 00185 Roma, Italy \\
15. Physics Department, Texas A\&M University, College Station, TX 77843, 
 USA \\
16. Dipartimento di Fisica Sperimentale dell'Universit\`a  di Torino and 
INFN, 10125 Torino, Italy \\
17. L.P.T.P, Faculty of Sciences, University Mohamed I, B.P. 524 Oujda, 
 Morocco \\
18. Dipartimento di Fisica dell'Universit\`a  di Roma Tre and INFN Sezione 
Roma Tre, 00146 Roma, Italy \\
19. Istituto per lo Sudio dello Spazio Interplanetario del CNR,
Sezione di Torino 10133 Torino, and INFN, 10125 Torino, Italy \\
$a$ Also Universit\`a  della Basilicata, 85100 Potenza, Italy \\
$b$ Also INFN Milano, 20133 Milano, Italy \\
$c$ Also IASF/CNR, Sezione di Bologna, 40129 Bologna, Italy \\
$d$ Also Universit\`a  di Trieste and INFN, 34100 Trieste, Italy \\
$e$ Also Dipartimento di Energetica, Universit\`a  di Roma, 00185 Roma, 
 Italy \\
$f$ Also Institute for Space Sciences, 76900 Bucharest, Romania \\
$g$ Macalester College, Dept. of Physics and Astr., St. Paul, MN 55105 \\
$h$ Also Department of Physics, Pennsylvania State University, University 
Park, PA 16801, USA \\
$i$ Also Dipartimento di Fisica dell'Universit\`a  della Calabria, Rende 
(Cosenza), Italy \\
$l$ Also Dipartimento di Scienze e Tecnologie Avanzate, Universit\`a  del 
Piemonte Orientale, Alessandria, Italy \\
$m$ Also U. Minn. Duluth Physics Dept., Duluth, MN 55812 \\
$n$ Also Dept. of Physics, MIT, Cambridge, MA 02139 \\
$o$ Also Intervideo Inc., Torrance CA 90505 USA \\
$p$ Also Resonance Photonics, Markham, Ontario, Canada\\
$q$ Also Dipartimento di Ingegneria dell'Innovazione dell'Universit`a di
Lecce and INFN, 73100 Lecce, Italy\\
\end{center}
\begin{frontmatter}
\clearpage
\begin{abstract}
The cosmic ray primary composition in the energy range between
$10^{15}$ and $10^{16}$ eV, i.e., around the ``knee" of the primary
spectrum, has been studied through the combined measurements of
the EAS-TOP air shower array (2005 m a.s.l., 10$^5$ m$^2$ collecting 
area) and the MACRO underground detector
(963 m a.s.l., 3100 m w.e. of minimum rock overburden, 920 m$^2$
effective area) at the National Gran Sasso Laboratories. The used
observables are the air shower size ($N_e$) measured by EAS-TOP and
the muon number ($N_{\mu}$) recorded by MACRO. The two detectors
are separated on average by 1200 m of rock, and located at a respective
zenith angle of about 30$^\circ$. The energy threshold at the surface
for muons reaching the MACRO depth is approximately 1.3 TeV. Such
muons are produced in the early stages of the shower development
and in a kinematic region quite different from the one relevant
for the usual $N_{\mu}-N_e$ studies. The measurement leads to a
primary composition becoming heavier at the knee of the primary
spectrum, the knee itself resulting from the steepening of the spectrum
of a primary light component (p, He). The result confirms the ones
reported from the observation of the low energy muons at the
surface (typically in the GeV energy range), showing that the
conclusions do not depend on the production region kinematics. Thus,
the hadronic interaction model used (CORSIKA/QGSJET) provides
consistent composition results from data related to secondaries
produced in a rapidity region exceeding the central one.
%(?)(is well consistent in representing the data).
% beyond the central rapidity region
%(?in realta' non c'e' una analisi di questo tipo) .
Such an evolution of the composition in the knee region
supports the ``standard" galactic acceleration/propagation models
that imply rigidity dependent breaks of the different components,
and therefore breaks occurring at lower energies in the spectra of
the light nuclei.
%which represents a main feature for the interpretation of the
%origin of galactic cosmic radiation (?? poco conclusivo? comunque
%aggiustare).
\end{abstract}
\end{frontmatter}
%\psdraft
\section{Introduction}
The study of the primary cosmic ray composition and of its evolution with
primary energy is the main tool in understanding the
cosmic ray acceleration processes. In particular the energy range
between $10^{15}$ and $10^{16}$ eV is characterized by breaks in
the size spectra of the different Extensive Air Shower (EAS)
components: electromagnetic (e.m.)~\cite{knee_em}, muon~\cite{knee_muon}, 
Cherenkov light~\cite{knee_cher}, and hadrons~\cite{knee_h}, which are 
therefore
interpreted as a break in the primary energy spectrum. It is
now recognized that the interpretation of such a feature could
provide a significant signature in understanding the galactic
cosmic radiation~\cite{peters}, \cite{zatsepin}, \cite{hillas},
\cite{bierman}, \cite{EW}.

Independent measurements based on the observation of the e.m. and
GeV muon components~\cite{kascade}, \cite{eastopcomp} lead to a
composition becoming heavier in this energy region. The situation
is more complex when other components are considered, thus showing
that further information is needed from independent observables (see,
e.g., \cite{sommers} and references therein). This is also useful
% and in particular, in the regions where spectral
%changes are observed, essentially all experimental results
%indicate a composition becoming heavier. A verification of this
%behavior is crucial in the interpretation of the observed Cosmic
%Ray spectrum, (c'e' dopo, riaggiustare?)
%Due to the low fluxes (I ~ 10^$^{-7}$ m${-2}$s$^{-1}$sr$^{-1}$)
%such energies have to be studied through indirect techniques, i.e.
%the Extensive Air Shower (EAS) one.
%The study of the primary composition in the EAS energy region
%requires however the use of independent observables in order
to cross check the information, reduce the dependence on the
hadron interaction model and particle propagation codes used, and
to have better control of fluctuations in shower development, and
therefore of event selection.

At the National Gran Sasso Laboratories, we have developed a program 
of systematic study of 
the surface shower size measurements from EAS-TOP
and the high energy muons ($E_{\mu}^{th} = 1.3 $ TeV) measured 
deep underground (MACRO).
Such muons originate from the decays of mesons produced
in the first interactions of the incident primary in the
atmosphere, and thus are from a quite different rapidity region than the GeV
muons usually used for such analyses ($x_{F}> 0.1$ or $ 0.2$, the
rapidity region being $y-y_{beam} \approx -(4.5 - 5.0) $ at $\sqrt
s \approx 1000$ TeV). The experiment provides therefore new data
related to the first stages of the shower development, from
secondaries produced beyond the central rapidity region.

EAS-TOP and MACRO operated in coincidence in their respective final
configurations for a live time of $\Delta T =$ 23,043 hours
between November 25, 1992 and May 8, 2000, 
corresponding to an exposure ${\Gamma \cdot \Delta T} \approx
4 \times 10^9$ m$^2$ s sr. We
present here an analysis of the full data set.  Further details
and partial results of the present work can be found in
\cite{macroeas90}, \cite{macroeas94} and \cite{icrc2001}.
%The measurement is of particular interest
%since it faces the cosmic ray primary composition problem in the
%energy range between $10^{15}$ and $10^{16}$ eV, i.e. the region
%characterized by a
%rather sharp (?)
%break in the size spectra of the different EAS components: e.m.
%\cite{knee_em}, muon \cite{knee_muon}, Cherenkov light
%\cite{knee_cher}, hadrons \cite{knee_h}, which is therefore
%interpreted as a break in the primary energy spectrum.
%Independent measurements (e.g. based on the observation of the
%e.m. and GeV muon components) lead to the conclusion of a
%composition becoming heavier in such energy region. The
%confirmation of such results through the observation of
%secondaries produced in a different kinematical region reduces the
%possible uncertainties connected to the interaction model used
%(giusto,... debole?? io toglierei le due frasi, Aurelio la
%seconda... ci vuole di piu' o di meglio?).
%The definition of the problem from the experimental point of view
%provides a fundamental element for the understanding of the origin
%of galactic cosmic radiation.
\section{The detectors}
The EAS-TOP array was located at Campo Imperatore (2005 m a.s.l.,
at about $30^\circ$ from the vertical at the underground
Gran Sasso Laboratories, corresponding to an atmospheric depth 
of 930 g cm$^{-2}$). Its e.m. detector (in which we are mainly
interested in the present analysis) was built of 35 scintillator
modules each 10 m$^2$ in area, resulting in a collecting area A $\approx$
10$^5$ m$^2$. The array was fully efficient for $N_e > 10^5$. In the
following analysis, we will use events with at least 7
neighboring detectors fired and a maximum particle density recorded
by an inner module (``internal events"). The EAS-TOP reconstruction
capabilities of the EAS parameters for such events are: ${{\Delta
N_e} \over N_e} \approx 10 \%$ above $N_e \approx 10^5$ for the
shower size, and $\Delta \theta \approx 0.9^\circ$ for the arrival
direction. The array and the reconstruction procedures are fully
described in~\cite{Aglietta93}.

MACRO, in the underground Gran Sasso Laboratories at 963 m a.s.l.,
with 3100 m w.e. of minimum rock overburden, was a large area
multi-purpose apparatus designed to
 detect penetrating cosmic radiation. 
A detailed
description of the apparatus can be found in~\cite{macro93}. In
this work we consider only muon tracks which have at least 4
aligned hits in both views of the horizontal streamer tube planes
out of the 10 layers composing the lower half of the detector, 
which had
dimensions $76.6 \times 12 \times 4.8$ m$^3$. 
The MACRO
standard reconstruction procedure~\cite{macro92} has been used,
which provides an accuracy due to instrumental uncertainties
and muon scattering in the rock of $0.95^\circ$ for the muon
arrival direction. The muon number is measured with an accuracy
$\Delta N_\mu < 1$ for multiplicities up to $N_\mu \approx 10$,
and $\Delta N_\mu \approx 1$ for $N_\mu > 10$; high multiplicity events
have been scanned by eye to avoid possible misinterpretations.
%($\Delta N_\mu$=2 will be the multiplicity bin used in the
%analysis, independent of the track density of the event.)

The two experiments are separated by a rock thickness ranging from
1100 to 1300 m, depending on the angle.
% (Fig. \ref{fig:perpi}).
The energy threshold at the surface for muons reaching the
MACRO depth ranges from $E_{\mu}^{th} = 1.3$ TeV to $E_{\mu}^{th}
= 1.8$ TeV within the effective area of EAS-TOP.\\
The two experiments operated with independent triggering conditions, set
to 1) four nearby detectors fired for EAS-TOP (corresponding to a
primary energy threshold of about 100 TeV), and 2) a single muon in
MACRO. Event coincidence is established off-line, using the
absolute time given by a GPS system with an accuracy better than 1
$\mu s$. The number of coincident events amounts to 28,160, of
which 3,752 are EAS-TOP ``internal events" (as defined above) and
have shower size $N_e > 2 \times 10^5$; among them 409 have $N_e >
10^{5.92}$, i.e., are above the knee observed at the corresponding
zenith angle~\cite{eastop99}. We present here the analysis of such
events, by using full simulations 1) of the detectors (based on
GEANT~\cite{GEANT}), 2) of the cascades in the atmosphere performed within the
same framework as for the surface data (CORSIKA/QGSJET
\cite{knapp98}), and 3) of the MUSIC code~\cite{Music97} for muon
transport in the rock. Independent analyses from the two experiments
separately are
reported in~\cite{macro97}, \cite{eastop99} and \cite{eastopcomp}.
%Further details and partial results of the present work can be found in
%\cite{macroeas90}, \cite{macroeas94} and \cite{icrc2001}.
%In \cite{icrc2001}
%different interaction models also implemented inside the CORSIKA
%code have been used, results being in qualitative agreement with
%the present ones.
\section{Analysis and results}
%The analysis technique has to be adapted to the specific trigger
%requirements for the detection of coincidence showers
%(both surface and underground detectors fired) with
%defined acceptance area for the EAS array (internal events), but
%undetermined a priori for the underground one, since MACRO
%detects muons also far from the shower axis defined inside the EAS
%array.
\subsection{The data}
\begin{figure}[htb]
\begin{center}
\includegraphics[width=12.0cm]{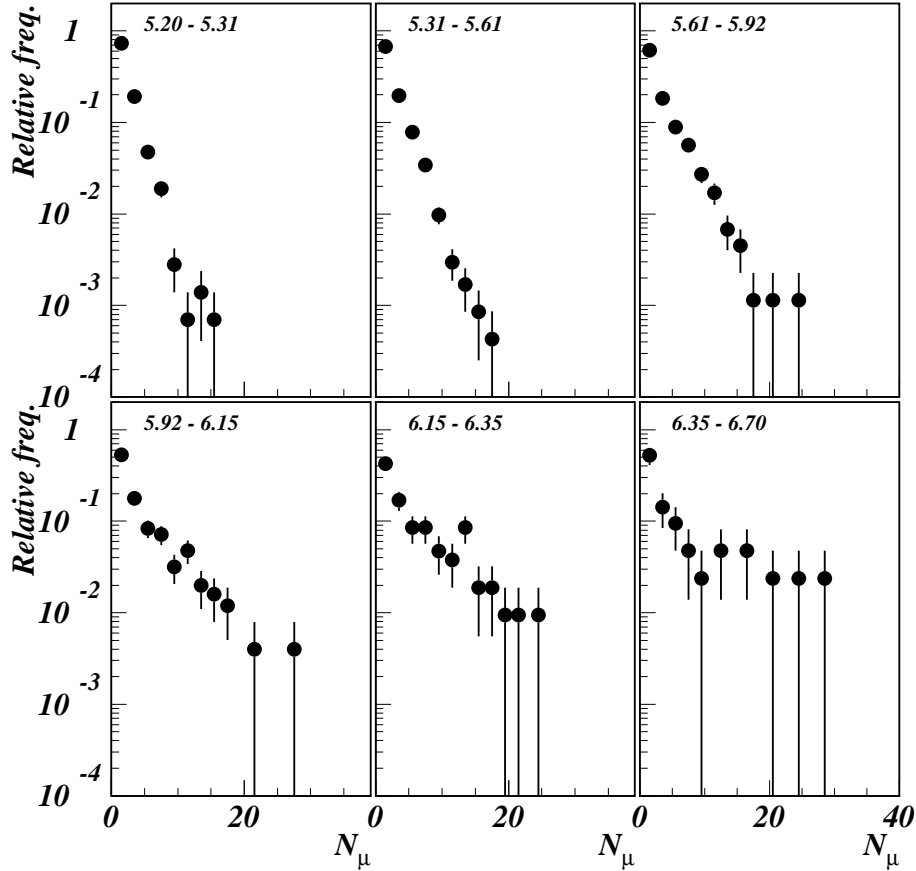}
\caption{Distributions of the relative frequencies
of the detected underground muon multiplicities in the 6 selected
size windows. See Tables
\protect\ref{tab:m1}--\protect\ref{tab:m6}. Notice the expected
increasing relative frequency of high multiplicity events as a
function of shower size. \label{fig:mulexp}}
\end{center}
\end{figure}

The experimental quantities considered are the muon
multiplicity distributions (for $N_{\mu} \ge 1$ as required by the
coincidence trigger condition) in several intervals of shower
sizes. We have chosen six intervals
of shower sizes covering the region of  the knee: \\
5.20 $< Log_{10}(N_e)\leq$ 5.31 (1432 events), 5.31 $<
Log_{10}(N_e)\leq$ 5.61 (2352 events), 5.61 $< Log_{10}(N_e)\leq$
5.92 (881 events), 5.92 $< Log_{10}(N_e)\leq$ 6.15 (252 events),
6.15 $< Log_{10}(N_e)\leq$ 6.35 (106 events) and 6.35 $<
Log_{10}(N_e)\leq$ 6.70 (42 events). The experimental relative
frequencies of the multiplicity distributions are shown in
Fig.\ref{fig:mulexp}.
%In order to make easier
For further analysis, the data have been grouped in variable
multiplicity bin sizes
%(the exact values of multiplicity bins and their content for the
%different size windows are
reported with their contents in Tables \ref{tab:m1}--\ref{tab:m6}.
\begin{center}
\begin{table}
\begin{tabular}{||c||c|c||c|c||c|c||c|c||}
\hline
\multicolumn{1}{||c||}{} & \multicolumn{2}{|c||}{Exp. data.} &
\multicolumn{2}{|c||}{ MC Light } & \multicolumn{2}{|c||}{ MC Heavy }
& \multicolumn{2}{|c||}{ Fit Light+Heavy } \\
\hline N$_\mu$ & f$_{ev}$ & $\sigma_{f_{ev}}~(\% )$ & f$_{ev}$ &
$\sigma_{f_{ev}}$ &
f$_{ev}$ & $\sigma_{f_{ev}}~(\% )$ & f$_{ev}$ & $\sigma_{f_{ev}}~(\% )$\\
\hline
\hline
 1--2  & 0.7353 & 3.1  & 0.7878 & 2.3    & 0.5656 & 3.2    & 0.7381 & 10.7 \\
 3--4  & 0.1927 & 6.0  & 0.1706 & 5.0    & 0.2436 & 4.9    & 0.1866 & 13.6 \\
 5--6  & 0.0475 & 12.2 & 0.0344 & 11.0   & 0.1206 & 7.0    & 0.0534 & 20.6 \\
 7--8  & 0.0189 & 19.0 & 0.0055 & 27.3   & 0.0509 & 10.8   & 0.0155 & 29.0 \\
 9--10 & 0.0028 & 50.0 & 0.0013 & 53.8   & 0.0135 & 20.7   & 0.0040 & 30.0 \\
11--12 & 0.0007 & 100.0& 0.0004 & 100.0  & 0.0053 & 34.0   & 0.0015 & 44.1 \\
13--14 & 0.0014 & 71.4 & 0.0    & 0.0    & 0.0006 & 100.0  & 0.0001 &100.0 \\
\hline
\end{tabular}
\caption{Relative frequency multiplicity distribution for size
window: $5.20 < Log_{10}(N_e) < 5.31$ (1432 events). The
simulated distributions are also reported for the Light, Heavy
components Monte Carlo (MC), and the resulting fit (see text). The number of
digits is chosen in order to show the one event level.
\label{tab:m1}}
\end{table}
\begin{table}
\begin{tabular}{||c||c|c||c|c||c|c||c|c||}
\hline
\multicolumn{1}{||c||}{} & \multicolumn{2}{|c||}{Exp. data.} &
\multicolumn{2}{|c||}{ MC Light } & \multicolumn{2}{|c||}{ MC Heavy } &
\multicolumn{2}{|c||}{ Fit Light+Heavy } \\
\hline N$_\mu$ & f$_{ev}$ & $\sigma_{f_{ev}}~(\% )$ & f$_{ev}$ &
$\sigma_{f_{ev}}$ &
f$_{ev}$ & $\sigma_{f_{ev}}~(\% )$ & f$_{ev}$ & $\sigma_{f_{ev}}~(\% )$ \\
\hline
\hline
 1--2  & 0.6743 & 2.5    & 0.7292 &  1.9   & 0.5147 & 2.7    & 0.6764 & 8.2   \\
 3--4  & 0.1973 & 4.7    & 0.1907 &  3.7   & 0.2010 & 4.2    & 0.1932 & 8.0   \\
 5--6  & 0.0782 & 7.5    & 0.0591 &  6.6   & 0.1472 & 5.0    & 0.0806 & 13.2  \\
 7--8  & 0.0344 & 11.0   & 0.0163 & 12.9   & 0.0836 & 6.6    & 0.0328 & 17.6  \\
 9--10 & 0.0098 & 20.4   & 0.0023 & 34.7   & 0.0374 & 9.9    & 0.0109 & 23.9  \\
11--12 & 0.0030 & 36.7   & 0.0016 & 37.5   & 0.0109 & 18.3   & 0.0038 & 21.1  \\
13--14 & 0.0017 & 52.9   & 0.0008 & 62.5   & 0.0036 & 30.6   & 0.0015 & 20.0  \\
15--16 & 0.0009 & 66.7   & 0.0    & 0.0    & 0.0007 & 71.4   & 0.0002 & 50.0  \\
17--18 & 0.0004 & 100.0  & 0.0    & 0.0    & 0.0    & 0.0    & 0.0    & 0.0\\
19--20 & 0.0    & 0.0    & 0.0    & 0.0    & 0.0007 & 71.4   & 0.0    & 0.0\\
\hline
\end{tabular}
\caption{As Table 1 for size  window: $ 5.31 < Log_{10}(N_e) <
5.61$ (2352 events). \label{tab:m2}}
\end{table}
\begin{table}
\begin{tabular}{||c||c|c||c|c||c|c||c|c||}
\hline
\multicolumn{1}{||c||}{} & \multicolumn{2}{|c||}{Exp. data.} &
\multicolumn{2}{|c||}{ MC Light } & \multicolumn{2}{|c||}{ MC Heavy } &
\multicolumn{2}{|c||}{ Fit Light+Heavy } \\
\hline N$_\mu$ & f$_{ev}$ & $\sigma_{f_{ev}}~(\% )$ & f$_{ev}$ &
$\sigma_{f_{ev}}$ &
f$_{ev}$ & $\sigma_{f_{ev}}~(\% )$ & f$_{ev}$ & $\sigma_{f_{ev}}~(\% )$ \\
\hline
\hline
 1--2  & 0.6118 & 4.3    & 0.6131 & 2.9    & 0.4412 & 4.1    & 0.5741 & 12.6  \\
 3--4  & 0.1839 & 7.9    & 0.2341 & 4.7    & 0.1689 & 6.7    & 0.2193 & 12.7  \\
 5--6  & 0.0885 & 11.3   & 0.0972 & 7.4    & 0.1214 & 7.9    & 0.1019 & 15.1  \\
 7--8  & 0.0568 & 14.1   & 0.0370 & 11.9   & 0.0988 & 8.7    & 0.0498 & 21.5  \\
 9--10 & 0.0272 & 20.6   & 0.0127 & 20.5   & 0.0799 & 9.8    & 0.0268 & 31.0  \\
11--12 & 0.0170 & 25,9   & 0.0053 & 32.0   & 0.0483 & 12.4   & 0.0143 & 35.0  \\
13--14 & 0.0068 & 41.2   & 0.0    & 0.0    & 0.0219 & 18.7   & 0.0046 & 50.0  \\
15--16 & 0.0045 & 60.0   & 0.0005 & 100.0  & 0.0106 & 26.4   & 0.0026 & 42.3  \\
17--18 & 0.0011 & 100.0  & 0.0    & 0.0    & 0.0030 & 50.0   & 0.0006 & 50.0 \\
19--20 & 0.0011 & 100.0  & 0.0    & 0.0    & 0.0030 & 50.0   & 0.0006 & 50.0 \\
21--28 & 0.0011 & 100.0  & 0.0    & 0.0    & 0.0030 & 50.0   & 0.0006 & 50.0 \\
\hline
\end{tabular}
\caption{As Table 1 for size  window: $ 5.61 < Log_{10}(N_e) <
5.92$ (881 events).
\label{tab:m3}}
\end{table}
\begin{table}
\begin{tabular}{||c||c|c||c|c||c|c||c|c||}
\hline
\multicolumn{1}{||c||}{} & \multicolumn{2}{|c||}{Exp. data.} &
\multicolumn{2}{|c||}{ MC Light } & \multicolumn{2}{|c||}{ MC Heavy } &
\multicolumn{2}{|c||}{ Fit Light+Heavy } \\
\hline N$_\mu$ & f$_{ev}$ & $\sigma_{f_{ev}}~(\% )$ & f$_{ev}$ &
$\sigma_{f_{ev}}$ &
f$_{ev}$ & $\sigma_{f_{ev}}~(\% )$ & f$_{ev}$ & $\sigma_{f_{ev}}~(\% )$ \\
\hline
\hline
 1--2  & 0.5318 &   8.6 & 0.4992 &   5.7 & 0.4353 &  7.0 & 0.4698 & 26.9 \\
 3--4  & 0.1786 &  14.9 & 0.2373 &   8.3 & 0.1315 & 12.8 & 0.1936 & 26.3 \\
 5--6  & 0.0833 &  21.8 & 0.1309 &  11.2 & 0.1013 & 14.6 & 0.1182 & 26.6 \\
 7--8  & 0.0714 &  23.5 & 0.0687 &  15.4 & 0.0927 & 15.2 & 0.0776 & 28.7 \\
 9--10 & 0.0318 &  35.2 & 0.0426 &  19.5 & 0.0754 & 17.0 & 0.0552 & 30.6 \\
11--12 & 0.0476 &  29.0 & 0.0147 &  33.3 & 0.0582 & 19.2 & 0.0317 & 37.5 \\
13--14 & 0.0198 &  44.9 & 0.0033 &  69.7 & 0.0409 & 23.0 & 0.0181 & 45.3 \\
15--16 & 0.0159 &  49.7 & 0.0016 & 100.0 & 0.0259 & 29.0 & 0.0112 & 45.5 \\
17--18 & 0.0119 &  58.0 & 0.0016 & 100.0 & 0.0172 & 35.5 & 0.0078 & 43.6 \\
19--24 & 0.0040 & 100.0 & 0.0000 &   0.0 & 0.0151 & 37.7 & 0.0060 & 50.0 \\
25--30 & 0.0040 & 100.0 & 0.0000 &   0.0 & 0.0065 & 56.9 & 0.0026 & 50.0 \\
\hline
\end{tabular}
\caption{As Table 1 for size  window: $ 5.92 < Log_{10}(N_e) <
6.15$ (252 events). \label{tab:m4}}
\end{table}
\begin{table}
\begin{tabular}{||c||c|c||c|c||c|c||c|c||}
\hline
\multicolumn{1}{||c||}{} & \multicolumn{2}{|c||}{Exp. data.} &
\multicolumn{2}{|c||}{ MC Light } & \multicolumn{2}{|c||}{ MC Heavy } &
\multicolumn{2}{|c||}{ Fit Light+Heavy } \\
\hline N$_\mu$ & f$_{ev}$ & $\sigma_{f_{ev}}~(\% )$ & f$_{ev}$ &
$\sigma_{f_{ev}}$ &
f$_{ev}$ & $\sigma_{f_{ev}}~(\% )$ & f$_{ev}$ & $\sigma_{f_{ev}}~(\% )$ \\
\hline
\hline
 1--2  & 0.4245 &  14.9 & 0.4271 &   8.9 & 0.3585 & 11.5 & 0.3818 & 39.5\\
 3--4  & 0.1698 &  23.6 & 0.1966 &  13.1 & 0.1557 & 17.4 & 0.1692 & 39.9\\
 5--6  & 0.0849 &  33.3 & 0.1492 &  15.1 & 0.0566 & 28.8 & 0.0857 & 48.2\\
 7--8  & 0.0849 &  33.3 & 0.1085 &  17.7 & 0.1085 & 20.8 & 0.1091 & 38.5\\
 9--10 & 0.0472 &  44.7 & 0.0475 &  26.7 & 0.0708 & 25.8 & 0.0639 & 37.4\\
11--12 & 0.0377 &  50.1 & 0.0475 &  26.7 & 0.0566 & 28.8 & 0.0541 & 37.9\\
13--14 & 0.0849 &  33.3 & 0.0102 &  57.8 & 0.0708 & 25.8 & 0.0523 & 40.0\\
15--16 & 0.0189 &  70.4 & 0.0068 &  70.6 & 0.0377 & 35.3 & 0.0283 & 39.6\\
17--18 & 0.0189 &  70.4 & 0.0000 &   0.0 & 0.0236 & 44.5 & 0.0164 & 42.1\\
19--20 & 0.0094 & 100.0 & 0.0000 &   0.0 & 0.0094 & 71.3 & 0.0066 & 42.4\\
21--22 & 0.0094 & 100.0 & 0.0034 & 100.0 & 0.0189 & 49.7 & 0.0142 & 39.4\\
23--26 & 0.0094 & 100.0 & 0.0034 & 100.0 & 0.0189 & 49.7 & 0.0142 & 39.4\\
27--30 & 0.0    &   0.0 & 0.0    &   0.0 & 0.0141 & 58.2 & 0.0098 & 41.8\\
\hline
\end{tabular}
\caption{As Table 1 for size  window: $ 6.15 < Log_{10}(N_e) <
6.35$ (106 events). \label{tab:m5}}
\end{table}
\begin{table}
\begin{tabular}{||c||c|c||c|c||c|c||c|c||}
\hline
\multicolumn{1}{||c||}{} & \multicolumn{2}{|c||}{Exp. data.} &
\multicolumn{2}{|c||}{ MC Light } & \multicolumn{2}{|c||}{ MC Heavy } &
\multicolumn{2}{|c||}{ Fit Light+Heavy } \\
\hline N$_\mu$ & f$_{ev}$ & $\sigma_{f_{ev}}~(\% )$ & f$_{ev}$ &
$\sigma_{f_{ev}}$ &
f$_{ev}$ & $\sigma_{f_{ev}}~(\% )$ & f$_{ev}$ & $\sigma_{f_{ev}}~(\% )$ \\
\hline
\hline
 1--2  & 0.5238 &  21.3 & 0.3712 &  10.9 & 0.3765 & 12.5 & 0.3752 & 57.5 \\
 3--4  & 0.1429 &  40.8 & 0.2096 &  14.5 & 0.1294 & 21.3 & 0.1486 & 64.7 \\
 5--6  & 0.0952 &  50.0 & 0.1441 &  17.4 & 0.0882 & 25.9 & 0.1016 & 64.9 \\
 7--8  & 0.0476 &  70.8 & 0.0699 &  25.0 & 0.0471 & 35.2 & 0.0526 & 62.9 \\
 9--10 & 0.0238 & 100.0 & 0.0611 &  26.7 & 0.0529 & 33.3 & 0.0549 & 59.4 \\
11--14 & 0.0476 &  70.8 & 0.1310 &  18.2 & 0.1353 & 20.8 & 0.1343 & 57.4 \\
15--18 & 0.0476 &  70.8 & 0.0087 &  71.3 & 0.0471 & 35.2 & 0.0379 & 56.2 \\
19--22 & 0.0238 & 100.0 & 0.0044 & 100.0 & 0.0412 & 37.9 & 0.0324 & 57.4 \\
23--26 & 0.0238 & 100.0 & 0.0000 &   0.0 & 0.0471 & 35.2 & 0.0358 & 59.5 \\
27--30 & 0.0238 & 100.0 & 0.0000 &   0.0 & 0.0353 & 40.8 & 0.0269 & 56.9 \\
\hline
\end{tabular}
\caption{As Table 1 for size  window: $ 6.35 < Log_{10}(N_e) <
6.70$ (42 events). \label{tab:m6}}
\end{table}
\end{center}

%The simulations have been performed in an energy range which
%includes the e.m. size values here considered (roughly between 100
%and 100000 TeV/particle) and in an angular range exceeding the
%aperture of the coincidence experiment. For the present analysis,
%the QGSJET \cite{qgsjet} hadronic interaction model implemented in
%CORSIKA has been used. In order to investigate the dependence of
%the results on the interaction model used, the simulation has been
%repeated, with reduced statistics, for other interfaces included
%in CORSIKA: DPMJET \cite{dpmjet}, HDPM \cite{hdpm} and SIBYLL
%\cite{sibyll}. The results are in general agreement, as discussed
%in \cite{icrc2001,taup2001}.
%We have generated 5 samples with different nuclear mass
%(proton, Helium, Nitrogen (CNO), Magnesium and Iron) with the
%same spectral index $\gamma = 2.62$ \cite{tech}. A number
%of events exceeding the experimental statistics has been simulated
%in all size bins. We have verified that in each of the considered
%size bithe fraction of each mass component contributing to that
%size is practically constant in energy. This is not straightforward
%sice, for a given mass, different energies contribute to the same size
%interval.
%\documentstyle[epsfig,psfig]{elsart}
%\begin{document}
\subsection{The simulation}
We have simulated atmospheric showers in an energy range which
includes the e.m. size values considered here (between 100 and
100,000 TeV/particle) and in an angular range exceeding the
aperture of the coincidence experiment. Shower simulations have been
performed with the QGSJET~\cite{qgsjet} hadronic interaction model
implemented in CORSIKA v5.62~\cite{knapp98} \footnote{Simulations,
using the other hadronic models in CORSIKA (DPMJET~\cite{dpmjet},
HDPM~\cite{hdpm} and SIBYLL~\cite{sibyll}) have been performed,
with reduced statistics, in order to verify the consistency
of the procedure. The results are discussed in~\cite{icrc2001,taup2001}.}.
Primary particles have been sampled in a solid  angle region of
the order of the area encompassing the surface array as seen from
the underground detector. The  solid angle corresponding to the
selected angular window is $\Omega = 0.0511$ ~sr. All muons with
energy $ E_{\mu} \ge$ 1 TeV reaching the surface have been
propagated through the rock down to the MACRO depth by means of
the  muon transport code MUSIC~\cite{Music97}; the accuracy of
this transport code has been verified by comparing its results to
those achieved with other Monte-Carlo simulations.
%"refined".
%The rock
%depth in the  region of  interest has  been corrected in all cases
%to  take into account  in a  parametric  form  the (small)
%differences due to a more   precise    description  of  the rock
%stratification, as obtained by  refined Monte Carlo simulation of
%muon transport (?).
Generated events having no muons surviving underground have been
discarded, while those having at least one surviving muon have been
folded with the underground detector simulation according to the
following  method, whose theoretical principles are discussed in
\cite{array}. We  have considered  an array of 39 (13 $\times$ 3) 
identical MACRO detectors adjacent to one another, covering an
area of 230.7 $\times$ 158.2 m$^2$. The shower axis is sampled over
the horizontal  area of the central  MACRO, and all hit detectors
are considered. For each hit  detector, the full GMACRO (GEANT
based) simulation of MACRO is invoked and is considered as a
different event. For each of these events, when considered at the
position of the ``real'' MACRO, the position of the shower core at
the surface is recalculated, the particle densities on EAS-TOP
counters are calculated and the trigger simulation is then
invoked. Particle densities are obtained from the lateral
distribution of the e.m. component of the shower as produced by
CORSIKA (with the analytical ``Nishimura-Kamata-Greisen'' option), 
taking into account the
fluctuations of the number of particles hitting the detector
modules and the full detectors' fluctuations~\cite{Aglietta93}.
%Such fluctuations are not poissonian, and  are  simulated after the
%parameterization of experimental  data. ADC  response, including
%the possible saturation, is also reproduced.
If the trigger threshold is reached, the reconstructions of both
EAS-TOP and MACRO are activated, thus producing results in the
same format as the real data. The resulting events, after combining
the simulated reconstructions of surface and underground detectors,
are eventually stored as simulated coincidence events.
%Following the strategy that we have been using in our previous preliminary
%works, in order to be able to reproduce a wide set of different
%composition models, we sample primaries from power spectra flatter
%than those of realistic models. Here we have generated
Five samples with different nuclear masses have been generated: 
proton, Helium, Nitrogen (CNO), Magnesium and Iron, all with the same 
spectral index $\gamma = 2.62$.
%For each of the considered size  bins, each mass component
%contributes with a different fraction. However, the fraction of a
%given mass component contributing to a given size window is found
%to be practically constant in the correspondent energy interval.
%Since the generation spectrum is different from the expected physical one,
%it is important that the analysis are performed in size bins small enough
%that the measured quantities are essentially energy independent. This has been
%verified for the muon multiplicities in the present analysis.
%(? dire meglio...)
Shower size bins have been chosen to be small enough so that
%to be independent from the
%spectral index used. In the present analysis this has been
%verified by Monte Carlo for underground muons:
no significant change in the shape of the muon multiplicity
distributions in each bin is observed for different, extreme
spectral indexes. A number of events exceeding the experimental
statistics have been simulated in each size bin.
%\begin{thebibliography}{99}
%\bibitem{knapp98}
%Knapp J. and Heck D., Extensive Air Shower Simulation with CORSIKA
%(Version 5.61), 1998.
%\bibitem{Music97}
%Antonioli P. et al., Astrop. Physics 7, 357, 1997.
%\bibitem{qgsjet} N.N.~Kalmikov ans S.S.~Ostapchenko,
%Yad. Fiz. {\bf 56} (1993) 105; Physics of Atomic Nuclei {\bf 58}
%(1995) 1728.
%\bibitem{array} G. Battistoni et al., Astrop. Phys. {\bf 7} (1995) 101.
%\end{thebibliography}
%\end{document}
%\Section{The analysis}
\subsection{The results}
The analysis is performed through independent fits of the
experimental muon multiplicity distributions in the selected
intervals of shower size. The simulated multiplicity distributions
have been used as theoretical expectations for the individual 
components, and the relative weights are the fit parameters.

The possibility that the experimental data could be reproduced
with a single mass component can be easily excluded for the
extreme (p or Fe) components, but also for medium mass primaries
(e.g., A = 14): the obtained values of ${\chi}^2$ are not satisfactory
(too large, the point will be addressed at the end of the
section). On the other hand, with a number of components larger than two
we cannot achieve better solutions, since in all cases the
minimization algorithm tends to force to zero the contribution of
an intermediate component.
% and unrealistic values of
%${\chi}^2$ (too small).
%??? dire meglio?
% However we could
%verify that the best results can be achieved when different masses
%are combined together with proper weights depending on the size
%bin. .
%Therefore we can exploit this property to infer the average mass
%composition as a function of energy. This is accomplished by
%fitting within each size bin the muon multiplicity distribution by
%means of a superposition of pure components.
%{\it Our statistics then implies that we can at best
%represent the primary beam as made up of two components.}
%!! 
This is mainly due to our limited statistics. For this reason we
performed our analysis by considering only two components in the
primary beam.
%!!
%at the same time.
We have tested two cases: a combination of p and
Fe components, and a combination of two admixtures: a ``Light''
($L$) and ``Heavy'' ($H$) one, built with equal fractions of p
plus He and Mg plus Fe, respectively.
%what we define as ``Light'' ($L$) and ``Heavy'' ($H$) admixtures,
%containing equal fractions of p plus He and Mg plus Fe
%respectively.
Preliminary results from the p+Fe analysis have been
presented in~\cite{icrc2001,taup2001}. Here we describe the final
analysis in terms of the $L$+$H$ admixtures.

The fit has been performed in the six quoted size windows by
minimizing the following expression for each
multiplicity distribution:
\begin{equation}
\xi^2 = \sum_i{\frac{(N^{exp}_i - p_L N^{L}_i - p_H N^{H}_i)^2 }{
\sigma_{i,exp}^2 + (p_L \sigma_{i,L})^2 + (p_H \sigma_{i,H})^2 } }
\end{equation}
where $N^{exp}_i$ is the number of events observed in the $i^{th}$
bin of multiplicity (with statistical uncertainty $\sigma _{i,exp}$), 
$N^{L}_i$ and $N^{H}_i$ are the numbers of
simulated events in the same $i^{th}$ multiplicity bin from the $L$
and $H$
%``Light'' and the ``Heavy''
components, respectively, $p_L$ and $p_H$ are the parameters (to
be fitted) defining the fraction of each mass component
contributing to  the same multiplicity bin, and $\sigma_{i,L}$ and
$\sigma_{i,H} $ are the statistical errors of the simulation. Such
an expression is close to that of a $\chi^2$, although, in
principle, it follows a different statistics, and in the following
we shall refer to it as if it were a genuine $\chi^2$.
%The fit results have been normalized to reproduce the
%observed number of coincident events in each size bin (see Tables
%\ref{tab:m1}--\ref{tab:m6}). In this way we have obtained the
%abundances of $L$ and $H$ nuclei as a function of the size and, as
%a consequence, the obtained size spectrum is by construction in
%agreement with the experimental one, for the same binning.
The values of the parameters $p_L$ and $p_H$ obtained from the
minimizations are given in Table \ref{tab:para}. The progressive
decrease of the ``Light" component in favor of the ``Heavy" one is
visible and significant at the level of 
2 standard deviations: the average $p_L$
value is 0.70 $\pm$ 0.04 below the observed knee in size (Log$_{10}$($N_e$)
= 5.92), and 0.28 $\pm$ 0.17 above.
By normalizing $p_L$ and $p_H$ to the observed number of
coincident events in each size bin (see Tables
\ref{tab:m1}--\ref{tab:m6}) we obtain the contribution to the
measured size spectrum of each component.

\begin{figure}[htb]
\begin{center}
%\begin{tabular}{cc}
\includegraphics[width=12.0cm]{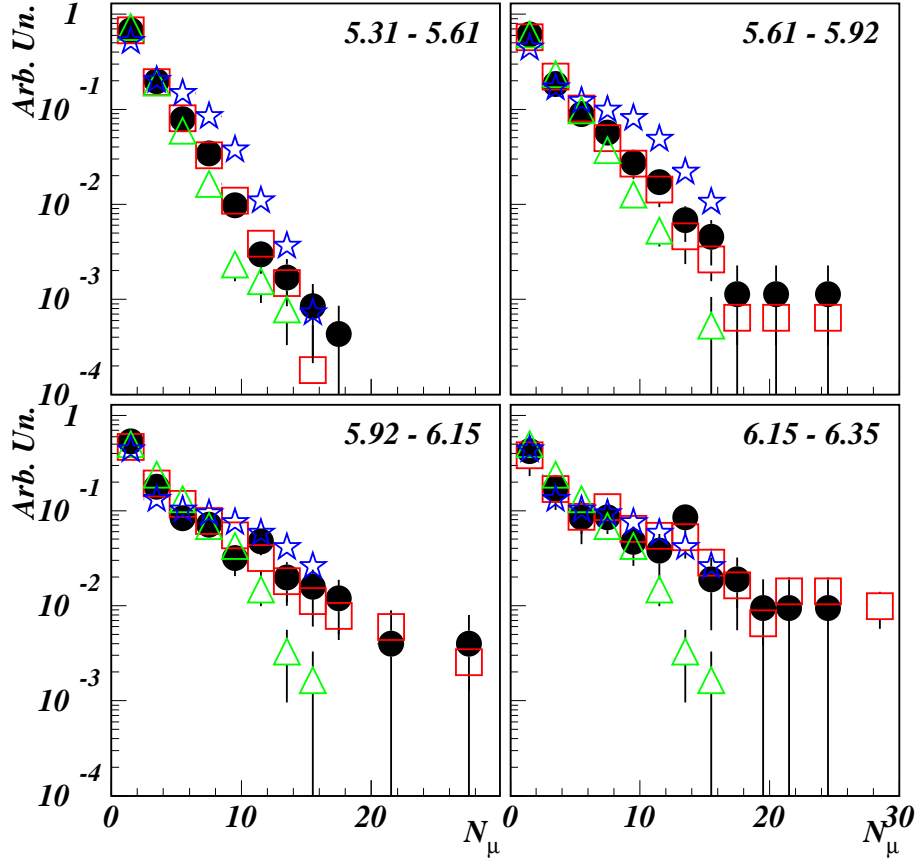} 
%\includegraphics[width=5.5cm]{lh3.eps} \\
%\includegraphics[width=5.5cm]{lh4.eps} &
%\includegraphics[width=5.5cm]{lh5.eps}
%\end{tabular}
\caption{Distributions of the relative frequencies
of the detected underground muon multiplicities (black points) together
  with the predictions for $L$ (open triangles) and $H$ (open stars)
admixtures in the QGSJET
  interaction model, and the ($L$+$H$) fit (open squares).
\label{fig:fits}}
\end{center}
\end{figure}

\begin{table}
\begin{center}
\begin{tabular}{|c|c|c|c|}
\hline
Log$_{10}$($N_e$) window & p$_L$ & p$_H$ & $\chi^2$/Nd.o.f. \\
\hline
5.20--5.31 & 0.74 $\pm$ 0.07 & 0.26 $\pm$ 0.11 & 5.5/5 \\
5.31--5.61 & 0.70 $\pm$ 0.05 & 0.30 $\pm$ 0.09 & 2.7/7 \\
5.61--5.92 & 0.66 $\pm$ 0.09 & 0.34 $\pm$ 0.14 & 11.4/9 \\
5.92--6.15 & 0.50 $\pm$ 0.17 & 0.50 $\pm$ 0.24 & 12.2/9 \\
6.15--6.35 & 0.30 $\pm$ 0.20 & 0.70 $\pm$ 0.32 & 4.7/10 \\
6.35--6.70 & 0.24 $\pm$ 0.32 & 0.76 $\pm$ 0.45 & 8.4/8 \\
\hline
\end{tabular}
\caption{ \label{tab:para} The fitted normalizations for the two
components (L, H) as a function of size (notice that the two
parameters are correlated, so that errors are not independent 
from one another).}
\end{center}
\end{table}
In Fig. \ref{fig:fits} the multiplicity distributions are shown for
the four most relevant size windows, together with the expected
$L$ and $H$ components, and their best fit combination.
%The progressive decrease of the $L$ component in favor of the $H$ one
%is visible (detto cosi' e' banale, comunque prima a commento della
%tabella; una figura questo non se la merita? si puo' dare un
%livello di fiducia a cui H aumenta e L diminuisce? si DOVREBBE!
%C'E' UN VOLONTARIO PER FARE L'ESERCIZIO? forse fare su figura
%finale in energia) .
%At this point it is of some interest to comment the shape of the
%multiplicity distribution. Both in the data and in the simulation
%it is quite clear that this

\begin{figure}[htb]
\begin{center}
%\begin{tabular}{cc}
%\includegraphics[width=5.5cm]{cno2.eps} &
%\includegraphics[width=5.5cm]{cno3.eps} \\
%\includegraphics[width=5.5cm]{cno4.eps} &
\includegraphics[width=12.cm]{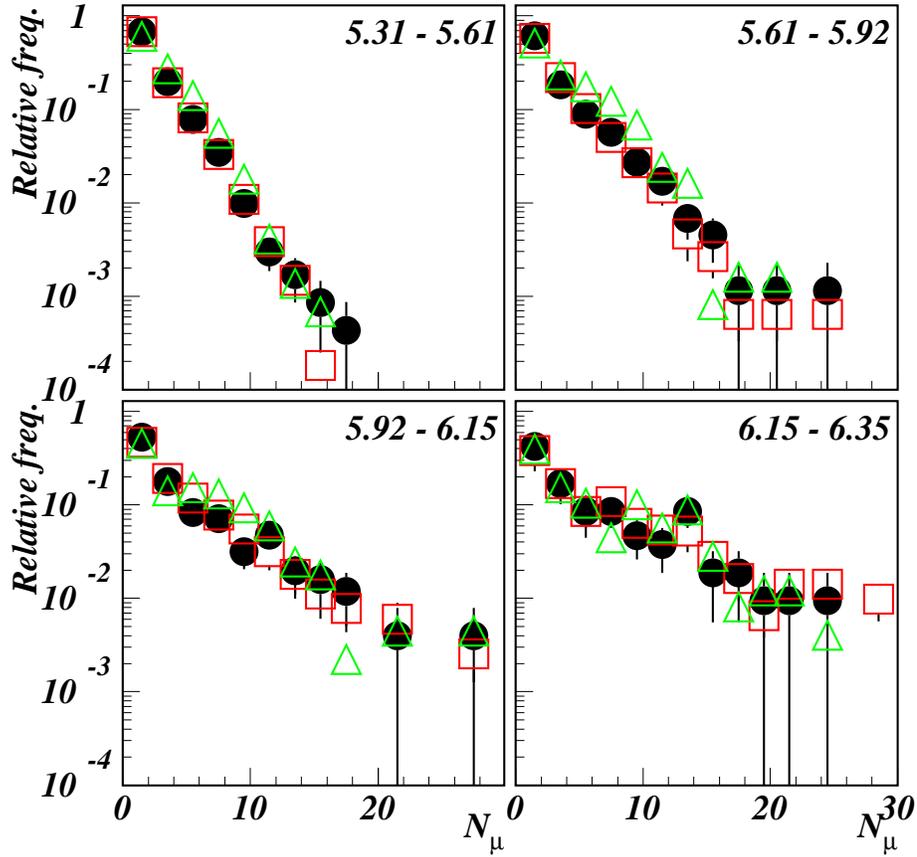}
%\end{tabular}
\caption{Distributions of the relative frequencies
of the detected underground muon multiplicities (black symbols)
together with the results of the fits for the QGSJET interaction
model ($L$+$H$) (open squares), compared with a fit with the CNO
component only (open triangles). \label{fig:fits3}}
\end{center}
\end{figure}

Regarding the shapes of the multiplicity distributions, it is
interesting to remark that they cannot be described by simple
single laws, and show some structure; this is evident in the
data and in the simulated Heavy components, however less so in the simulated
Light ones. The origin of such structure is entirely geometric and
due to the interplay between the typical size of muon bundles with
the two length scales of the MACRO detector. Small bundle
sizes can be entirely contained in the detector while,
when the size increases, this becomes impossible along the width of
the detector. Bundles of even larger size exceed also the length of MACRO.
This fact is well taken into account by the simulation, and in
fact the fit reproduces correctly this change of structure,
which is typical of large bundles (i.e., high energies and large
masses). 
%In fact these remarks are confirmed 
The effect is evident when comparing with a
single component fit, say the CNO group that has an intermediate
average atomic number. The results of the fit are presented in
Fig. \ref{fig:fits3}. CNO primaries alone provide good fits in the
higher size bins (due to the limited statistics), but below and
just above the knee at $Log_{10}(N_e)=5.92$, the large $\chi^2$ 
values indicate the
failure to reproduce the shape of the multiplicity distribution
(see Table \ref{tab:cnof}).

\begin{table}
\begin{center}
\begin{tabular}{|c|c|}
\hline
Log$_{10}$($N_e$) window & $\chi^2$/Nd.o.f. \\
\hline
5.20--5.31 & 17.3/6 \\
5.31--5.61 & 49.9/8 \\
5.61--5.92 & 45.6/10 \\
5.92--6.15 & 16.8/10 \\
6.15--6.35 &  4.7/11 \\
6.35--6.70 &  8.7/9  \\
\hline
\end{tabular}
\caption{ \label{tab:cnof} The $\chi^2$ values resulting from the
fits to the CNO (A=14) component alone, as a function of shower
size. }
\end{center}
\end{table}

\subsection{Interpretation of the data}

For a given size window, the contribution of each primary mass
group derives from a different energy region: the higher the mass
number, the higher the corresponding energy.
The size-energy mapping far from shower maximum is model dependent,
and in our analysis is based on
CORSIKA/QGSJET. From the full simulation chain we also calculate
the probabilities $\epsilon_\alpha (E, \Delta_i N_e)$
 for a primary belonging to mass group $\alpha$ ($\alpha~=~L,~H$)
and of energy $E$ to give a coincident event in the $i^{th}$ size window
$\Delta_i N_e$.
To evaluate the average mass composition we use a logarithmic energy binning (3 bins per energy decade),
starting from 100 TeV/nucleus. From the simulation we obtain the
number of events ($n^\alpha_j (\Delta_i N_e)$) that a primary of mass
group $\alpha$ will produce in the $j^{th}$ energy bin, when the
detected
size is in the windows $\Delta_i N_e$. Therefore the total number
of events that the primary mass group $\alpha$ produces in the size
window $\Delta_i N_e$ is the sum of $n^\alpha_j (\Delta_i N_e)$ over
the energy bins.

%Considering two mass groups, $A$ and $B$, w
We require that the number of experimentally observed
events in the size window $\Delta_i N_e$ be equal to:

\begin{equation}
N^{exp}(\Delta_i N_e) = p^L(\Delta_i N_e) \sum_j {n^L_j (\Delta_i
N_e)} + p^H(\Delta_i N_e) \sum_j {n^H_j (\Delta_i N_e)}
\label{eq:eq1}
\end{equation}

where $p^L$ and $p^H$ are the fit coefficients for the given size
window $\Delta_i N_e$. These are normalized, so that $p^L = 1 -
p^H$ in each size window.
This leaves an overall renormalization factor $K$ free
in order to satisfy eq.\ref{eq:eq1}, so we obtain the renormalized
quantities $n^{*\alpha}_j~=~K~n^{\alpha}_j$.
The corrected estimated number of primaries of mass
group $\alpha$ for each size window belonging to energy bin $j$
can thus be obtained by applying the efficiencies $\epsilon_{\alpha}
(E_j, \Delta_i N_e)$:
\begin{equation}
m^{\alpha}_j (\Delta_i N_e) = p^\alpha(\Delta_i N_e) {n^{* \alpha}_j
(\Delta_i
N_e)/\epsilon_{\alpha} (E_j,\Delta_i N_e)} \label{eq:eq2}
\end{equation}
%%with a similar expression for mass group $B$.$
Then, since the $j^{th}$ energy bin may receive contributions from
different size windows, we have to sum over $i$ (the size window
index):
\begin{equation}
M^{\alpha}_j   = \sum_i m^{\alpha}_j (\Delta_i N_e)  = \sum_i
{p^\alpha(\Delta_i N_e) {n^{* \alpha}_j (\Delta_i
N_e)/\epsilon_{\alpha} (E_j,\Delta_i N_e)}} \label{eq:eq3}
\end{equation}

\begin{figure}[htb]
\begin{center}
\includegraphics[width=8.cm]{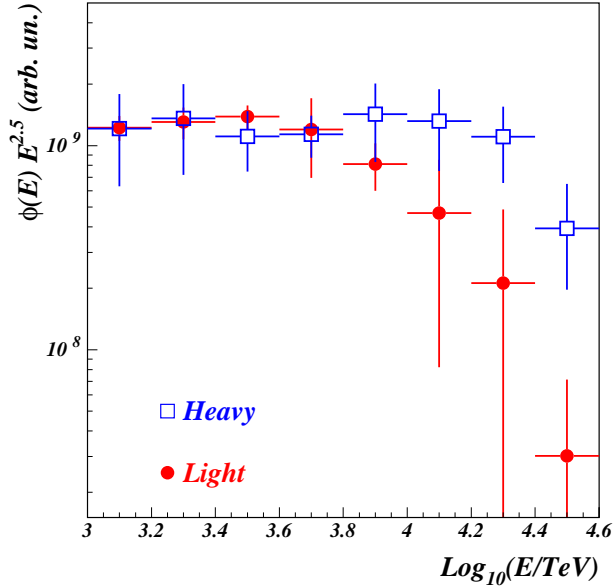}
\caption{Energy spectra estimates for $L$ and $H$  admixtures.
\label{fig:spettrie}}
\end{center}
\end{figure}

$M^{L}_j$ and $M^{H}_j$ provide estimates of the energy spectra of the $L$ and $H$ mass groups,
%as a function of energy,
%Our result for Light and Heavy is
presented in Fig. \ref{fig:spettrie}. There we plot the spectra starting 
from $10^3$ TeV since with our selection of size, this is the energy at 
which the heaviest component has reached a
significant triggering efficiency.

%{\it 
%Within our limited statistics
A steepening by about $\Delta \gamma = 0.7 \pm 0.4$ of the light 
mass group spectrum just at the knee ($\sim$4$\times$10$^{15}$ eV) is
observed,  
%{\it 
assuming  power law behaviors crossing at the knee position.
%We wish to remark that
Although these distributions cannot be used to obtain a
direct representation of the actual cosmic ray spectrum,
due to the two mass groups schematization
and the choices of their components,
%. The study of our simulation show that the parameters
%of these spectra depend on the choice of the original spectrum
%used to sample events and, to a lesser extent, on the composition
%of the mass group. On the other hand,
the relative proportion of ``Light'' and ``Heavy'' admixtures
turns out to be quite stable with respect to the mentioned
parameters;
within this approximation, the resulting all-particle spectrum 
would show a $\Delta \gamma = 0.4 \pm 0.1$.

\begin{figure}[htb]
\begin{center}
\includegraphics[width=12.cm]{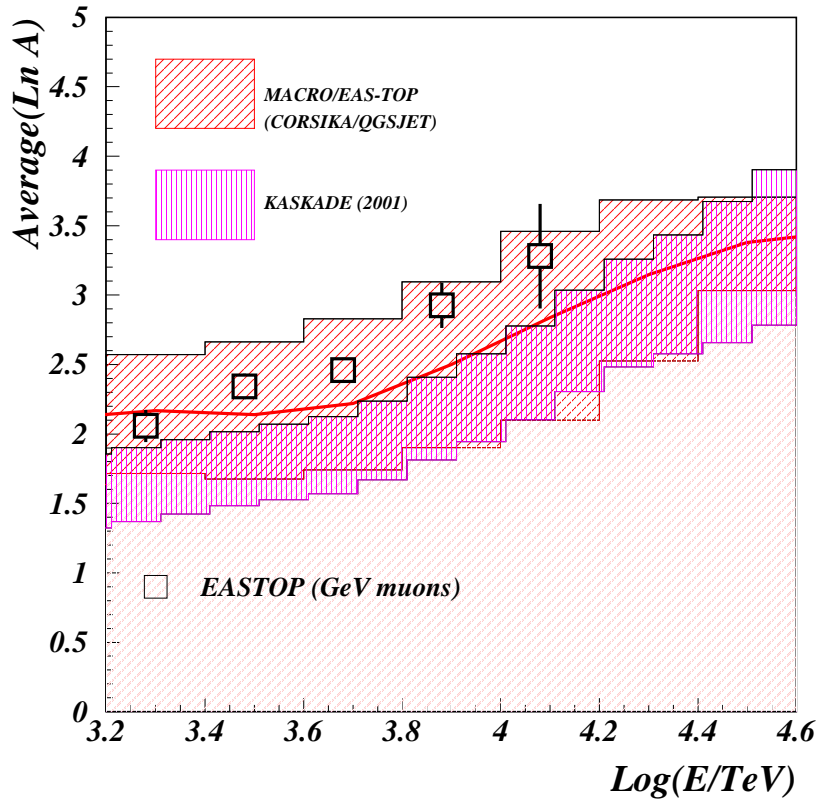}
\caption{$<ln A>$ vs primary energy (continuous line). The hatched areas
represent the 68\% uncertainty range due to the
statistical error. We also superimpose
the results of KASCADE \protect\cite{kascade}
and of EAS-TOP~\cite{eastopcomp} (open squares).
\label{fig:aval}}
\end{center}
\end{figure}

We make use of the values of $M^{L}_j$ and
$M^{H}_j$ so obtained to compute the mean value of the natural logarithm of the
primary mass ($<\ln A>$) as a function of energy:
\begin{equation}
<\ln A(E_j)> = \frac{\ln A^L~ M^L_j + \ln A^H~ M^L_j}{M^L_j +
M^H_j} \label{eq:eq7}
\end{equation}
%where we have used the proper values of Light and Heavy admixtures,
with $ln A^L = 0.5(ln A^p + ln A^{He})$, and 
$ln A^H = 0.5(ln A^{Mg} + ln A^{Fe})$. 
The uncertainty on $<\ln A(E_j)>$ has been
obtained by propagating the uncertainties on the fit coefficients.
The result is reported in Fig.~\ref{fig:aval} together with 
the results of KASCADE~\cite{kascade} and EAS-TOP alone~\cite{eastopcomp}, 
where these analyses has been performed using e.m. size
and GeV muons detected at surface level.
The good agreement shows that the results do not depend
on the selected muon energy.
The $<\ln A(E_j)>$ obtained by MACRO alone~\cite{macro97}, on the
basis of the HEMAS Monte Carlo code~\cite{hemas}, has a milder 
energy dependence and appears to
be in contrast with those presented here above $Log_{10}(E)> 4.2$. In our
opinion this is due to a weakness of the HEMAS model,
based on parameterizations of UA5 results~\cite{UA5}. The possible
shortcomings of the HEMAS model were already discussed in~\cite{isv98,isv98_2}.

The allowed region for $<ln A(E)>$ obtained from our analysis is also
consistent with the theoretical expectations from refs.
\cite{EW},~\cite{bez}, and~\cite{swordy}.
%In Fig.~\ref{fig:aval2}, where
%we compare our results to the theoretical expectations from refs.
%\cite{EW}, \cite{bez}, and \cite{swordy} (
%continuous lines). 
%
%\begin{figure}[htb]
%\begin{center}
%\includegraphics[width=12.cm]{macroeas_only.eps}
%\caption{$<ln A>$ vs primary energy as obtained from this work
%compared to the 
%theoretical expectations (\cite{EW}, \cite{bez}, and
%\cite{swordy}). \label{fig:aval2}}
%\end{center}
%\end{figure}

\section{Conclusions}
The analysis N$_e$-N$_\mu^{TeV}$ events collected by the
MACRO/EAS-TOP Collaboration
%in about 10 years of data taking
at the Gran Sasso Laboratories
%favors the conclusion of
points to a primary composition becoming heavier around the knee
of the primary spectrum (i.e., in the energy region
$10^{15}-10^{16}$ eV). The result is in good agreement with the
measurements of other experiments based on the observation of the
e.m. and muon components at ground level. The muon energies
detected in the present experiment are however about three orders
of magnitude larger than in previous N$_e$-N$_\mu$ experiments, and
therefore the parent pions are produced in a different kinematic
region (at the edges of the fragmentation region, rather than in
the central one) and in the first stages of the cascade
development. A good overall consistency of the interaction model used
(CORSIKA/QGSJET) in describing the yield of secondaries over a
wide rapidity region is thus obtained\footnote{An earlier
experiment performed in coincidence between a surface EAS array
and a deep underground detector in 1970's reached a different
conclusion, possibly due, in our opinion, to the smaller
dimensions of the underground detector~\cite{KGF}}.
%The present measurement reinforce the hypothesis that the observed knee in the
%spectrum is due to the break of the spectrum of a light component,
%that has to be interpreted
The present data explain therefore the observed knee in the cosmic
ray primary spectrum as due to the steepening of the spectrum of a
light component (p, He) at $E_0 \approx 4 \times 10^{15}$ eV, of
$\Delta\gamma = 0.7 \pm 0.4$. Such an effect can be interpreted in the
``standard" framework of the acceleration/propagation processes of
galactic cosmic radiation that predict, as a general feature,
rigidity dependent breaks for the different nuclei, and therefore
appearing at lower energies for the lighter ones.

\section*{Acknowledgements}
We gratefully acknowledge the support of the director and of the staff of the 
Laboratori Nazionali del Gran Sasso and the invaluable assistance of the 
technical staff of the Institutions participating in the experiment. We thank 
the Istituto Nazionale di Fisica Nucleare (INFN), the U.S. Department of 
Energy and the U.S. National Science Foundation for their generous support 
of the MACRO experiment. We thank INFN, ICTP (Trieste), WorldLab and NATO 
for providing fellowships and grants (FAI) for non Italian citizens.

\end{document}